\documentstyle[11pt,newpasp,twoside,epsf,epsfig]{article}
\def\edcomment#1{\iffalse\marginpar{\raggedright\sl#1\/}\else\relax\fi}
\reversemarginpar

\begin{document}
\title{Flux ratio [\ion{Ne}{v}] 14.3/24.3 as a test of collision strengths}
\author{Rubin, R. H.}
\affil{NASA Ames Research Center, Moffett Field, CA 94035--1000, USA;
rubin@cygnus.arc.nasa.gov}

\begin{abstract}
	From ISO [\ion{Ne}{v}] 14.3/24.3~$\mu$m line flux ratios,
we find that 10 out of 20 planetary nebulae (PNs) have measured ratios 
below the low-electron density ($N_e$) 
theoretical predicted limit. 
Such astronomical data serve to provide important tests of atomic data, 
collision strengths in this case. 
In principle, well-calibrated measurements of the [\ion{Ne}{v}] 
14.3/24.3 flux ratio could improve upon the existing atomic data.
\end{abstract}

\section{Introduction}

In an earlier study of 
PNs with Infrared Space Observatory
(ISO) observations, 
Rubin et~al.\ (2001)
found evidence that the useful 
$N_e$ 
diagnostic line flux ratio
[\ion{Ne}{v}] F(14.3~$\mu$m)/F(24.3~$\mu$m) 
was out of range of theoretical predictions
using current atomic data.
In particular, NGC~6818 with a measured flux ratio of 0.71 was significantly
out of bounds in the low-$N_e$ limit. 
Thus $N_e$ could not be derived; we concluded that perhaps the calculations 
of the effective collision strengths for these lines should be revisited. 
Now we include all PNs in the ISO archive that have both lines well measured. 

\section{Results, discussion, and conclusions}

Table 1 presents our results.
The last column has the derived $N_e$ or an
$*$ when the flux ratio is out of bounds in the low-$N_e$ limit 
(see figure~5 and table~6 in Rubin et~al.\ 2001) based on the
effective collision strength from Lennon \& Burke (1994).
Using more recent values from
Griffin \& Badnell (2000), the discrepancy is somewhat less but
still present.

\noindent
$\bullet$
Ten out of 20 PNs and 15 of the 28 PN ISO/TDTs (TDT \# identifies 
an observation)
in the Table yield 
[\ion{Ne}{v}]  F(14.3)/F(24.3) 
below the low-$N_e$ theoretical limit.
While there might be systematic errors affecting the ISO data, we note that
the observed aperture sizes match for this line pair. 
The low-$N_e$ limit is governed by the effective collision strengths.
We believe this work points to the need to reevaluate the effective
collision strengths for this ion.

\noindent
$\bullet$
Even when the line flux ratio is within ``legal bounds",
as is the case for the
3 symbiotic stars (HM~Sge, RR~Tel, and V~1016~Cyg)
included in the Table,
and a value for $N_e$ may be obtained,
if the collision strengths are to be revised as 
indicated by much of the data here, then these derived
$N_e$ values are likely to require a revision upward.

\newpage

\begin{figure}[top]
\vskip-0.1truein
\centerline{
\psfig{figure=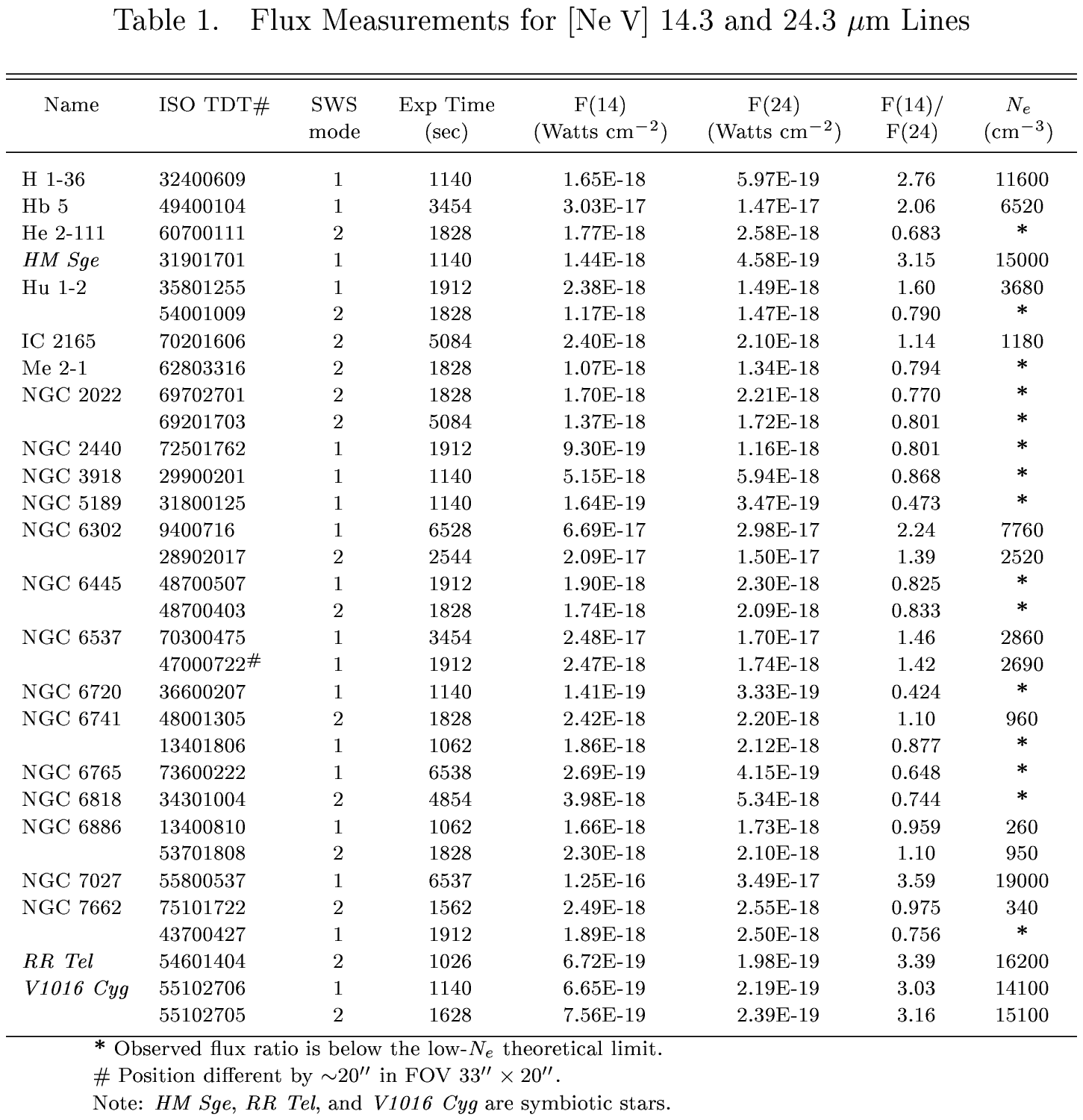,width=1.1\textwidth}}
\end{figure}

\acknowledgments{RHR acknowledges support from the NASA Long-Term Space 
Astrophysics (LTSA) program.}


\begin{references}
\reference Griffin, D.C., \& Badnell, N.R.\ 2000, J. Phys. B, 33, 4389
\reference Lennon, D.J., \& Burke, V.M. 1994, A\&AS, 103, 273
\reference Rubin, R.H., Dufour, R.J., Geballe, T.R., Colgan, S.W.J.,
Harrington, J.P., Lord, S.D, Liao, A.L., \& Levine, D.A. 2001,
in Spectroscopic Challenges of Photoionized Plasmas, ASP Conference 
series, Vol.\ 247, Eds. G.J. Ferland \& D.W. Savin, p.\ 479
(astro-ph/0109398)
\end{references}
\end{document}